# A good and computationally efficient polynomial approximation to the Maier-Saupe nematic free energy

Ezequiel R. Soule[1] and Alejandro D. Rey[2]

1. *Institute of Materials Science and Technology (INTEMA), University of Mar del Plata and National Research Council (CONICET), J. B. Justo 4302, 7600 Mar del Plata, Argentina*

2. *Department of Chemical Engineering, McGill University, Montreal, Quebec H3A 2B2, Canada*

*Date: April 16, 2010*

A new computational strategy is proposed to approximate, with a simple but accurate expression, the Maier-Saupe free energy for nematic order. Instead of the traditional approach of expanding the free energy with a truncated Taylor series, we employ a least-squares fitting to obtain the coefficients of a polynomial expression. Both methods are compared, and the fitting with at most five polynomial terms is shown to provide a satisfactory fitting, and to give much more accurate results than the traditional Taylor expansion. We perform the analysis in terms of the tensor order parameter, so the results are valid in uniaxial and biaxial states.

Liquid crystal (LCs) materials display intermediate degrees of positional and orientational order, between crystalline solids and liquids [1-2]; the simplest LC phase is the nematic phase that displays only orientational order. To describe the state of order in a nematic phase, the second moment of the orientation distribution function is usually sufficient, in the sense that most of the relevant experimental information (anchoring, textures, defects) is captured by it. This symmetric and traceless quadrupolar tensor is known as the tensor order parameter **Q**, and can be written as

$$\mathbf{Q} = S\left(\mathbf{nn} - \frac{\boldsymbol{\delta}}{3}\right) + \frac{P}{3}(\mathbf{ll} - \mathbf{mm}) \tag{1}$$

where $S$ is the scalar uniaxial order parameter, $P$ is the biaxial order parameter, $\boldsymbol{\delta}$ is the identity matrix and **n**, **l** and **m** are the eigenvectors **Q** [1]. The scalar uniaxial order parameter $S$ measures the degree of molecular alignment along the average orientation **n**, and plays a central role in investigating phase transitions, phase separation, pattern formation and multi-phase equilibria in the mixtures of polymer/monomers and LCs [2]. The biaxial order parameter measures the deviation of the molecular alignment distribution from axial symmetry, it plays a fundamental role in the formation of defects, interfaces, texturing, and biaxial states.

In computational modeling, the accuracy and usefulness of predictions depends both on the model and on the numerical methods. A simple expression for the bulk nematic free energy, is the Landau-de Gennes theory (LdG) [1-8], in terms of the invariants of the order parameter. Usually, a fourth-order polynomial is used:

$$f_{LdG} = a\mathbf{I}_2 + b\mathbf{I}_3 + c\mathbf{I}_4 + ... = a\mathbf{I}_2 + b\mathbf{I}_3 + c\mathbf{I}_2^{\ 2} + ... \tag{2}$$

where $f_{LdG}$ represents the dimensionless free energy density, $\mathbf{I}_i$ are $i^{\text{th}}$-order invariants of **Q**. According to Caley-Hamilton theorem, $\mathbf{I}_2=\mathbf{Q:Q}$ and $\mathbf{I}_3=(\mathbf{Q.Q}):\mathbf{Q}$ are the only independent invariants of **Q** [6] and all higher-order invariants are written in terms of these two. LdG model is based in the phenomenological Landau theory, which is based in the general assumption that the free energy in the vicinity of a phase transition can be written as an analytical function of some phase variable that describes the state of the system. In LdG model for nematic liquid crystals, this phase variable is the nematic order parameter. The coefficients *a, b, c…* are phenomenological parameters that must be measured; for thermotropics *a* is a function of temperature. Generalizations of the model to mixtures, and to account for the presence of external fields and surfaces are available [2].

Another very popular model is the Maier-Saupe theory (MS) and its extensions and modifications [2, 3, 8-15, which is based in statistical thermodynamics and has no phenomenological parameters:

$$f_{MS} = \frac{3}{4}\Gamma \mathbf{Q}{:}\mathbf{Q} - \ln\left(\int_0^{2\pi}\int_0^{\pi} \exp\left(\frac{3}{2}\Gamma \mathbf{Q}{:}\left(-\mathbf{u}\mathbf{u}\frac{\boldsymbol{\delta}}{3}\right)\right)\sin\theta\, d\theta\, d\varphi\right) \qquad (3)$$

where $\Gamma = 4.54 T/T_{NI}$ is the nematic interaction parameter, $T$ is the absolute temperature, $T_{NI}$ is the nematic-isotropic first-order transition temperature, **u** is the molecular unit vector and the integration is over the unit sphere. The first term arises form orientation-dependent energetic interaction, and the last term is the logarithm of the partition function [9-12]. This expression for $f_{MS}$ is a function of the invariants of **Q**. This is a mean-field theory, which assumes that the nematic-isotropic transition is produced by attractive interactions, and excluded-volume effects are neglected

From a computational point of view, a transient multidimensional model based on LdG is much more attractive because it requires short calculation times and standard computational resources, as opposed to MS that requires the numerical solution of an integral for each time step and space node. Also, the LdG is a phenomenological theory whose parameters can be fitted from experimental data, so that the behaviour of a given system could be represented in a wide range of conditions (using the adequate sets of parameters), whereas the accuracy of MS theory is solely determined by the applicability of its assumptions to the experimental system under consideration. But the fact that the parameters in LdG theory are phenomenological and the parametric data is not always available is a disadvantage. This disadvantage is particularly important for mixtures involving LCs [2], where experimental data would be required for every mixture composition analyzed. In these cases it is necessary to use a model with no phenomenological parameters, like the MS theory. A usual strategy that allows to keep the computational simplicity of LdG, but with no adjustable parameters, consist in using a polynomial expression obtained from a Taylor series expansion of MS free energy[11,16,17]. This can be easily done for uniaxial nematics by calculating the derivatives of the free energy as a function of $S$ analytically [3,16,17].

With this strategy, a simple polynomial expression with no adjustable parameters becomes available. The problem is that the Taylor expansion is a very poor approximation to the MS free energy. Katriel et al [11] have shown that this series only converges to the exact solution in a limited range of values of the order parameter. Das and Rey [3] explicitly compared phase diagrams for mixtures of a LC with an isotropic solvent (polymer), predicted by a modification of MS (including an excluded volume term in $\Gamma$) and by a LdG expression based on a fourth-order Taylor series expansion for uniaxial nematics ($P=0$). They found that the Taylor expansion was very unaccurate, when compared to the exact solution of Maier – Saupe free energy obtained by an accurate numerical solution of the integral. As an example the value of $\Gamma$ at which nematic-isotropic transition is predicted [3] with the Taylor expansion is 4.315, with a value of $S = 0.77$, while the value according to MS is 4.54, with $S = 0.44$. In addition, in some cases the phase diagrams predicted by the Taylor expansion were not only quantitatively but also qualitatively different to the ones predicted by using the accurate solution of $f_{MS}$. If we take into account that MS theory does not always represents well the experimental data, an unaccurate approximation to MS theory can lead to significant differences between theoretical predictions and experiment.

In this communication we generalize the polynomial approximation to $f_{MS}$ in order to account for the biaxial nematic state, and we propose a new simple strategy that provides an excellent approximation to $f_{MS}$, keeping the simple polynomial expression. The computational strategy consists in calculating the polynomial coefficients not from a Taylor series, but from a least-squares fitting of the polynomial to the exact values of $f_{MS}$. Emphasis is put on taking into account biaxiality, since LdG models are specially used to simulate defect cores and flow-induced orientation [3].

We wish to point out that there are previous works that analyze and compare the Maier Saupe and Landau-de Gennes approaches in terms of physics involved, predictive capability and limitations of each theory (see for example refs 2, 11 and 12). The present work's objective is to analyze the computational efficiency of the models and the accuracy of two different strategies to approximate the full solution of MS theory, and not to compare the theories from a physical or a formal point of view.

As mentioned above, the free energy is a function of the invariants of **Q**, and these invariants are only a function of $S$ and $P$ (more specifically, $\mathbf{I}_i$ is a linear combination of the products $S^j P^k$ with $i=j+k$). Moreover, as can be seen in equation (3), $f_{MS}$ is composed by a polynomial term and a non-polynomial term which is a function of $\Gamma\mathbf{Q}$, so we can rewrite the polynomial expansion as:

$$f_{LdG-MS} = \frac{3}{4}\Gamma\mathbf{I}_2(S,P) + \sum_{i=2}^{N} b_i \mathbf{I}_i(\Gamma S, \Gamma P) \tag{4}$$

where $N$ is the order of the highest-order polynomial term. The fitting was performed obtaining the coefficients $b_i$ that minimize the sum of $(f_{MS} - f_{LdG-MS})^2$ calculated at several values of $\Gamma S$ and $\Gamma P$, spanning different ranges. In order to better represent the value of the transition temperature and the value of $S$ at the transition, the free energy for $P = 0$ was weighted with a factor of 10 when performing the fitting. It is noted that in order to perform the fitting, the Maier-Saupe free energy was computed by solving the integral numerically, using Romberg algorithm [18] with a small tolerance value to ensure high accuracy.

The Taylor expansion could be obtained by calculating the derivatives of $f_{MS}$ with respect to $\Gamma S$ and $\Gamma P$, but the direct analytical derivation and evaluation of these derivatives is not trivial. Instead, an indirect method was devised and implemented; we take into account the form of equation 4, which indicates that the free energy is expressed as a function of the invariants of $\mathbf{Q}$, and the invariant of order $i$ involves all the polynomial terms of order $i$. Consequently, all the $i^{\text{th}}$-order derivatives of the free energy are related through the invariants and the expansion coefficients $b_i$. The Taylor expansion for the uniaxial case ($P = 0$) can be easily done (and it has been done, see for example Das and Rey [3]). Knowing the coefficients for the uniaxial case and taking into account that $\mathbf{I}_2 = 2/3S^2 + 2P^2$ and $\mathbf{I}_3 = 2/9S^3 - 2SP^2$, the coefficients $b_i$ can be calculated and the whole expression constructed. For example, the third-order term of the Taylor series for the uniaxial case is $1/105*(\Gamma S)^3$ [3,16], and according to equation 4, this term is equal to $b_3\mathbf{I}_3$ with $P = 0$, so $1/105 = 2/9*b_3$.

Figures 1a, 1b and 1c show the free energy density from the MS model, the fourth-order Taylor expansion and a fourth-order least-squares fitting, at different values of $\Gamma$. It can be seen that in general the approximation of the Taylor expansion is poor, neither the values of $S$ at the local minimum $S>0$, nor the shape of the curve, are well represented. The accuracy of the fitting is considerably better.

The values of the fitting coefficients, and $\Gamma$ and $S$ at the transition are shown in Table 1 for different cases. This table also shows the maximum values of $\Gamma S$ and $\Gamma P$ used in each case (the fitting was always performed for $S$ and $P >0$), as well as the standard error, $\sigma$, calculated as:

$$\sigma = \sqrt{\frac{\sum_{i=1}^{n}\left[\ln\left(\int_0^{2\pi}\int_0^{\pi}\exp\left(\frac{3}{2}\Gamma\mathbf{Q}:\left(\boldsymbol{\sigma\sigma}-\frac{\boldsymbol{\delta}}{3}\right)\right)\sin\theta d\theta d\varphi\right)-\sum_{i=2}^{N}b_i\mathbf{I}_i(\Gamma S,\Gamma P)\right]^2}{n}} \quad (5)$$

where $N$ is the total number of points. Nine equidistant values of $S$ and four equidistant values of $P$ have been used to perform the fitting and calculate $\sigma$ in all cases. From the maximum value of $\Gamma S$, a maximum value of $\Gamma$ was calculated (and included in the table) considering that the equilibrium value of $S$ is a function of $\Gamma$. This means that if the coefficients are used for a value of $\Gamma$ higher than this limit, the value of $\Gamma S$ calculated at equilibrium will be outside the range of the fitting.

The accuracy of the approximation for all the sets of coefficients reported in Table 1 is good when used in the indicated range; if used outside these ranges the accuracy becomes lower and even unphysical results (like $S>1$) can be obtained. It can be seen that the two least accurate approximations are the one plotted in figure 1 (which is still a good approximation), and the one corresponding to the broadest range of S and P used. The fitting is always much more accurate than the Taylor expansion, for example, the standard error for the fourth-order Taylor expansion (calculated in the same way that for the fittings) in the range $\Gamma S < 2.8$, $\Gamma P < .6$ is 0.43, as opposed to .0033 for the fourth-order fitting. In some cases, the fitting was performed in similar ranges using both four and five terms, for comparison. It can be seen, by comparing the values of $\sigma$, $\Gamma_{NI}$ and $S_{NI}$, that when using a fifth order term the fitting is much improved, specially when the range of P is larger. As the values of $\Gamma$ or P increase, the accuracy of the fitting with a given number of polynomial terms decreases, so more terms are necessary.

The optimal set of coefficients will depend on the conditions that are being simulated, that will determine the relevant ranges of $\Gamma S$ and $\Gamma P$. A broad range of values is covered in Table 1, but if the relevant values for a specific simulation were outside the ones reported, then a new fitting should be performed, probably using a higher-order polynomial.

As a representative application, figure 2 shows a comparison of the equilibrium order parameter S as a function of dimensionless temperature ($\Gamma^{-1}$), calculated with the numerical solution, the Taylor expansion and the fitting using four term in both cases. Again, the values calculated with Taylor expansion are far from the exact values, while the approximation provided with the fitting is much better.

It is worth noting that the equilibrium state predicted by Maier – Saupe free energy in the absence of external fields, flow or surfaces is always uniaxial (P=0). Considering any of these effects would imply the addition of extra terms to the free energy, that can lead to biaxial stationary points. For example, in the precence of surfaces, the inclusion of gradient terms in the free energy density can lead to a stationary solution including topological defects, and in the vicinity of these defects a biaxial state can be found. Defects can also be generated during the dynamic evolution of the system towards the equilibrium, and dynamics simulations are usually performed using a fourth-order LdG free energy [19,20]

These results are not restricted to pure LC and can be used when modeling LC solutions. The MS theory for a binary mixture, where the volume fraction of liquid crystal is $\phi$, is written as [3,14,15]:

$$f_{MS} = \frac{3}{4}\Gamma^2\phi^2\mathbf{Q}:\mathbf{Q} - \phi\ln\left(\int_0^{2\pi}\int_0^{\pi}\exp\left(\frac{3}{2}\Gamma\phi\mathbf{Q}:\left(\boldsymbol{\sigma\sigma} - \frac{\boldsymbol{\delta}}{3}\right)\right)\sin\theta d\theta d\varphi\right) \quad (6)$$

$$f_{LdG-MS} = \frac{3}{4}\Gamma\phi^2\mathbf{I}_2(S,P) + \phi\sum_{i=2}^{N}b_i\mathbf{I}_i(\Gamma\phi S, \Gamma\phi P) \quad (7)$$

As the summation, which is the relevant term for the fitting, is the same than before except that the variables change from $\Gamma S$ and $\Gamma P$ to $\Gamma\phi S$ and $\Gamma\phi P$, the same coefficients $b_i$ reported in Table 1 are to be used, but the polynomial is expressed in terms of $\Gamma\phi S$ and $\Gamma\phi P$, and the ranges reported in Table 1 are ranges of $\Gamma\phi S$ and $\Gamma\phi P$.

In conclusion, a new strategy for approximating the Maier-Saupe free energy with a polynomial (Landau-de Gennes) expression was devised and implemented with least-squares fitting of the numerical solution of the non-analytical part of free energy. Biaxiality, important in defect cores and in the presence of external fields was taken into full account. The new approach was compared with the usual fourth-order Taylor expansion strategy. It has been shown that the fitting performs much better than the Taylor expansion, allowing a very accurate approximation using a very simple, computationally convenient, expression. In this way, the simplicity and convenience of Landau-de Gennes theory can be conserved, but accuracy is not sacrificed as in the case with the Taylor expansion. It has been shown that these results can also be used for LC solutions and blends. Finally, this strategy is not restricted to Maier-Saupe free energy, in principle the same procedure could be applied to any other phase transition theory, provided that the free energy can be evaluated (analytically or numerically) as a function of the relevant order parameters.

**Acknowledgements**

ADR acknowledges partial support from the Natural Science an Engineering Research Council of Canada (NSERC), the Petroleum Research Fund. ERS acknowledges support from the National Research Council of Argentina (CONICET).

**Legends to the Figures**

**Figure 1.** Free energy calculated solving numerically the integral in Maier Saupe expression (full), with a fourth order Taylor expansion (dash), and with a fourth order least-square fitting (dot) for: (a) $\Gamma = 4.315$, (b) $\Gamma = 4.54$, (c) $\Gamma = 5$

**Figure 2.** Equilibrium value of $S$ as a function of $\Gamma^{-1}$ (dimensionless temperature), calculated solving numerically the integral in Maier Saupe expression (full), with a fourth order Taylor expansion (dash), and with a fourth order least-square fitting (dot)

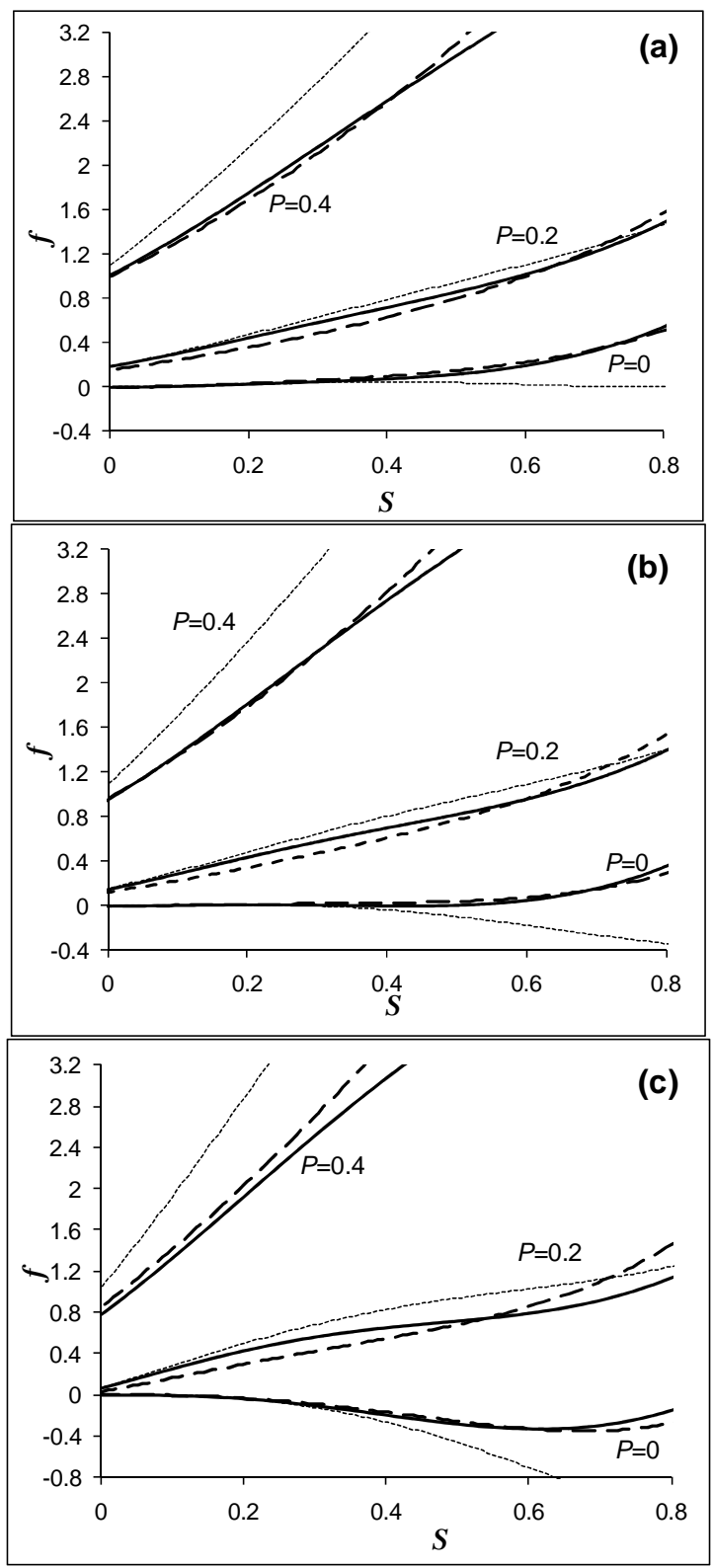

**Figure 1**

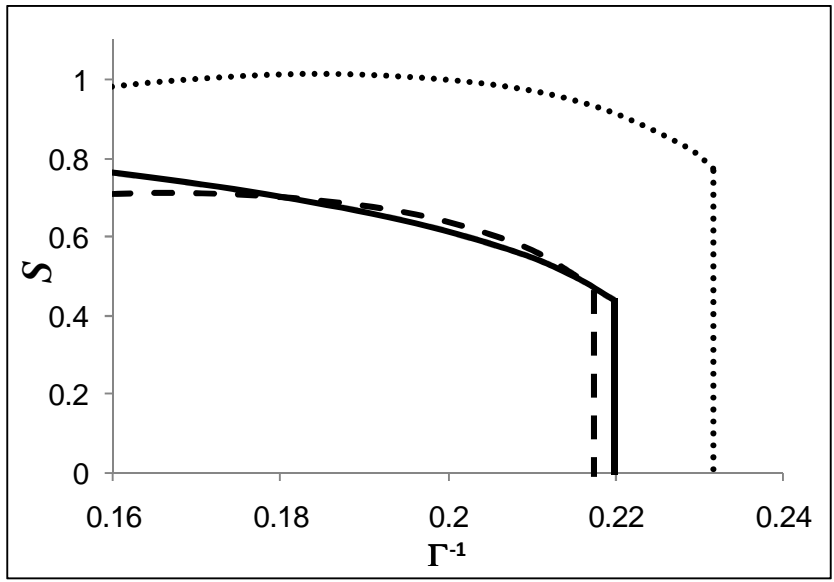

Figure 2

**Table 1.** Maier-Saupe free energy density fitting.

| $\Gamma S_{max}$ | $\Gamma P_{max}$ | $\Gamma_{max}$ | $b_2$ | $10^2 b_3$ | $10^3 b_4$ | $10^3 b_5$ | $10^2 \sigma$ | $\Gamma_{NI}$ | $S_{NI}$ |
|---|---|---|---|---|---|---|---|---|---|
| 2.8 | 0.6 | 4.8 | 0.154 | 3.11 | -3.81 | - | 0.33 | 4.559 | 0.456 |
| 2.8 | 1 | 4.8 | 0.151 | 3.01 | -3.05 | - | 0.62 | 4.576 | 0.452 |
| 2.8 | 1 | 4.8 | 0.148 | 4.14 | -2.37 | -2.44 | 0.15 | 4.539 | 0.452 |
| 2.8 | 2 | 4.8 | 0.149 | 3.66 | -2.45 | -1.34 | 0.46 | 4.561 | 0.491 |
| 3.5 | 0.6 | 5.26 | 0.156 | 2.75 | -3.74 | - | 0.42 | 4.555 | 0.404 |
| *__3.5__* | *__1__* | *__5.26__* | *__0.154__* | *__2.61__* | *__-3.11__* | *__-__* | *__1.19__* | *__4.609__* | *__0.46__* |
| 3.5 | 1 | 5.26 | 0.147 | 3.85 | -2.03 | -1.88 | 0.32 | 4.57 | 0.496 |
| 3.5 | 2 | 5.26 | 0.147 | 3.67 | -2.31 | -1.40 | 0.96 | 4.581 | 0.499 |
| 4.2 | 0.5 | 5.85 | 0.158 | 2.34 | -3.35 | - | 0.57 | 4.56 | 0.351 |
| 4 | 0.6 | 5.71 | 0.151 | 3.01 | -2.47 | -0.878 | 0.64 | 4.587 | 0.491 |
| 4.25 | 1 | 5.92 | 0.147 | 3.33 | -1.86 | -1.24 | 0.71 | 4.622 | 0.519 |
| *__4.5__* | *__0__* | *__6.1__* | *__.154__* | *__3.06__* | *__-3.85__* | *__-__* | *__.301__* | *__4.58__* | *__.437__* |
| 4.5 | 2 | 6.1 | 0.143 | 3.52 | -1.63 | -1.33 | 1.74 | 4.655 | 0.576 |

The coefficients in the fifth row (highlighted), are used in figure 1; the coefficients in the second-to-last rows (highlighted) are used in figure 2. The dash in the column corresponding to $b_5$ implies that for those cases four terms were used.